\documentclass[aps,prb,reprint,superscriptaddress,showpacs]{revtex4-1}
\usepackage{bm}
\usepackage{graphicx}
\begin{document}

\title{Level-crossing spectroscopy of nitrogen-vacancy centers in diamond:
sensitive detection of paramagnetic defect centers}

\author{S.~V.~Anishchik}
\email[]{svan@kinetics.nsc.ru} \affiliation{Voevodsky Institute of
Chemical Kinetics and Combustion SB RAS, 630090, Novosibirsk,
Russia}

\author{V.~G.~Vins}
\affiliation{VinsDiam Ltd., Russkaya str., 43, 630058,
Novosibirsk, Russia}

\author{K.~L.~Ivanov}
\affiliation{International Tomography Center SB RAS, 630090,
Novosibirsk, Russia} \affiliation{Novosibirsk State University,
630090, Novosibirsk, Russia}

\begin{abstract}
We report a magnetic field dependence of fluorescence of diamond
single  crystals containing NV$^-$ centers. In such spectra,
numerous sharp lines are found, which correspond to Level
Anti-Crossings (LACs) in coupled spins systems comprising an
NV$^-$ center. Theoretical modeling of such ``LAC-spectra'' enables
characterization of paramagnetic defect centers and determination
of their magnetic resonance parameters, such as zero-field splitting and
hyperfine coupling constants. The outlined method thus enables
sensitive detection of paramagnetic impurities in diamond crystals.

\end{abstract}


\maketitle

\section{Introduction}

The negatively charged nitrogen-vacancy center (NV$^-$ center) in
diamond signle crystals is of great  interest due to its unique
physical properties.\cite{Doherty2013} NV$^-$ centers represent a
promising molecular system for many applications as well as an
excellent test system for single-molecule spectroscopy, quantum
information
processing\cite{Gruber1997,Wrachtrup2001,Jelezko2004r,Childress2006,
Wrachtrup2006,Hanson2006b,Gaebel2006,Santori2006o,Waldermann2007,Maurer2012,
vanderSar2012,Dolde2013,Dolde2014,Pfaff2014} and nanoscale
magnetometry
\cite{Taylor2008,Balasubramanian2008,Maze2008,Rittweger2009,
Acosta2009,Fang2013}.

The ground state of the NV$^-$ centers is a triplet state. The
triplet  ground state is split (due to the electron dipole-dipole interaction) and the eigen-states have different energies depending on the spin projection on the symmetry axis.
The energy term lowest in energy is the term with zero projection,
$M_S$, of the electron spin on the molecular axis. The splitting
between this term and the terms with the projections $+1$ and $-1$ at
zero magnetic field is $D=2.87$~GHz. For symmetry reasons (the
system has $C_{3v}$ symmetry) the $\pm 1$ terms are degenerate (due to axial symmetry of the diolar tensor).

Upon light-induced transitions to the excited state of the NV$^-$ center, the evolution of the system depends on the electron spin state. The reason is the spin projection-selective inter-system crossing from the excited triplet state to the excited singlet state. For the
$M_S=0$ state this process is inefficient, whereas for the
$M_S=\pm 1$ states the excited singlet state is formed with a
relatively high yield. The singlet excited state eventually decays
to the ground state due to inter-system crossing; again, the rate
of this process depends on $M_S$ being the highest for $M_S=0$.
Consequently, after a few excitation cycles the $M_S=0$ state is
enriched; in other words, strong non-equilibrium electron spin
polarization the NV$^-$ center is formed. This effect is usually
referred to as optically-induced spin polarization
\cite{Loubser1977,Manson2006,Delaney2010,Goldman2015} despite the
fact that in this state the average projection of the electron spin is zero for any direction in space (no macroscopic net polarization is formed). Optically
generated polarization of NV$^-$ centers is of significance for
many applications.

One of the methods to study the properties of NV$^-$ centers and
their  interaction with other defect centers is provided by the
analysis of the photo-luminescence intensity of the NV$^-$ centers
and its magnetic field dependence.
\cite{VanOort1989,Epstein2005,Hanson2006,Rogers2008,Rogers2009,Lai2009,Armstrong2010,
Anishchik2015} The external magnetic field changes the spin
polarization value: this effect can be monitored by a reduction of
the luminescence intensity of NV$^-$ centers at particular
magnetic field strengths. \cite{Manson2006} Typically, the
magnetic field dependence of the photo-luminescence contains a
smooth background with sharp lines on top of it. For observing
such lines it is necessary to orient precisely the diamond
crystal: these lines can only be detected when the magnetic field
vector is parallel to the [111] axis of the diamond crystal
lattice. Even a slight misalignment strongly broadens the lines
and reduces their magnitude such that the sharp lines vanish.
These lines are attributed to Level Anti-Crossings (LACs) either
in the NV$^-$ center or in an extended spin system comprising an
NV$^-$ center and another paramagnetic defect center in diamond.
The most prominent line is observed at 1024 G, where there is an
LAC of the triplet levels of the ground state of the NV$^-$
center. Other lines are referred to, perhaps, misleadingly, as
cross-relaxation lines. \cite{VanOort1989} However, we clearly
demonstrate in this work that all sharp lines are coming from the
coherent spin dynamics at specific LACs. Thus, it is reasonable to
name the sharp lines ``LAC-lines'' and the corresponding magnetic
field dependencies can be named ``LAC-spectra''. By an LAC (also
termed ``avoided crossing'') we mean the following situation. Let
us imagine that at a particular field strength a pair of levels,
corresponding to quantum states $|K\rangle$ and $|L\rangle$, tends
to cross, i.e., to become degenerate. However, when there is a
perturbation matrix element $V_{KL}\neq 0$, which mixes the
levels, the degeneracy is lifted with a consequence that the
crossing is avoided. Importantly, at an LAC efficient coherent
exchange of populations of the $|K\rangle$ and $|L\rangle$ states
occurs,
\cite{Colegrove1959,Pravdivtsev2013,Pravdivtsev2013a,Clevenson2016},
i.e., LAC efficiently mediate spin polarization transfer.

In this work, we perform an experimental and theoretical study  of
LAC-spectra of diamond single crystals containing NV$^-$ centers.
For observation of LAC-lines we make an efficient use of lock-in
detection; this method dramatically increases the sensitivity.
Thus, we detect the luminescence in the presence of a
small-amplitude modulation of the external magnetic field. We
demonstrate that the LAC-lines are prominent at small modulation
frequencies. By optimizing the experimental settings we can
strongly enhance weak LAC-lines and reveal many new LAC-lines,
which were previously unknown. We propose a simple and efficient
theoretical approach to calculating LAC-spectra. A comparison
between theory and experimental data enables indirect detection of
paramagnetic centers and precise determination of their magnetic
parameters.

\section{Experimental results}

\subsection{Materials and methods}
Details of the experimental method are outlined in a previous
publication \cite{Anishchik2015}.

Experiments were performed using two samples, here named  SMP1 and
SMP2 of synthetic diamond single crystals. The two samples differ
in the content and concentration of paramagnetic impurities. The
samples were grown at high temperature and high pressure in the
Fe-Ni-C system. As-grown, the crystals were iradiated by fast
electrons of the energy of 3~MeV, the irradiation dose was
$10^{18}$~el/cm$^2$; after that the samples were annealed in
vacuum for two hours at a temperature of 800$\rm ^o$. The average
concentration of the NV$^-$ centers for the SMP1 and SMP2 samples
are $1.4\times10^{18}$ and $9.3\times10^{17}$~cm$^{-3}$,
respectively; the concentration of the so-called P1 centers (see
explanation below) is 23 and 55~ppm, respectively. Hence, the SMP2
sample contains more paramagnetic impurities than the SMP1 sample.
Both samples contain carbon and nitrogen atoms in natural
abundance.

The sample was located in an external magnetic field,  which is a
sum of a permanent field $B_0$ and an oscillating field of a small
amplitude $B_m$:
\begin{equation}
    B=B_0+B_m \sin(2\pi f_m t). \label{bmod}
\end{equation}
Here $f_m$ is the modulation frequency.  The sample  was
irradiated by light of a wavelength of 532~nm and power of 400~mW.
The beam direction was either parallel or perpendicular to the
magnetic field vector \textbf{B}$_0$. In all experiments we worked
with linearly polarized light and varied the orientation of the
electric polarization vector \textbf{E} with respect to the
\textbf{B}$_0$ vector. Most experiments were done with \textbf{E}
parallel to \textbf{B}$_0$ and with \textbf{E} perpendicular to
\textbf{B}$_0$. The luminescence intensity was measured by a
photo-multiplier; the modulated luminescence signal was an input
signal for the lock-in amplifier. The modulation frequency $f_m$
was varied in a range from 10~Hz to 100~kHz. In all experiments
presented here the modulation amplitude $B_m$ was 0.5~G. All
experiments were performed at room temperature.

\begin{figure}
   \includegraphics[width=0.42\textwidth]{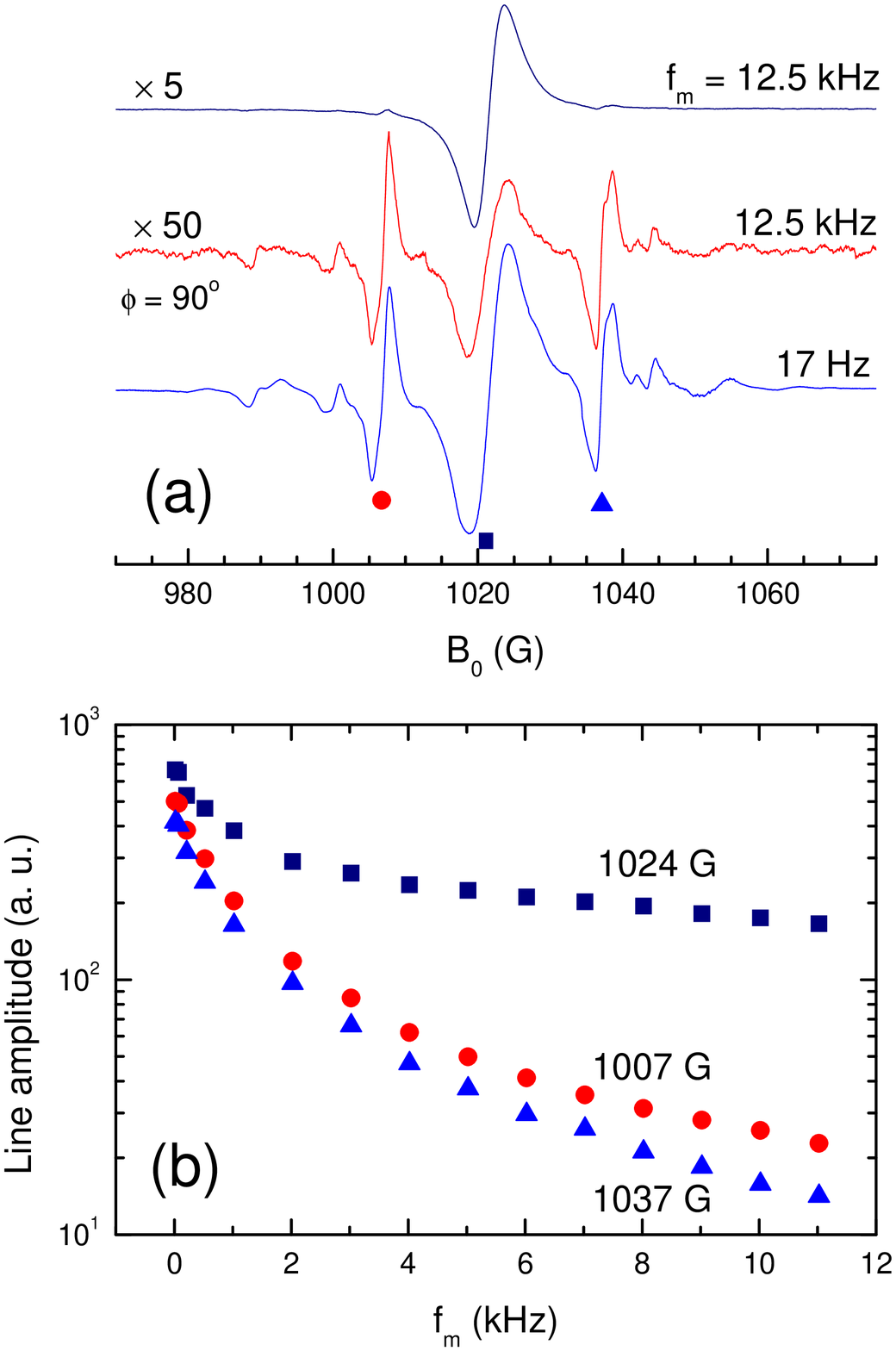}
   \caption{(a) Experimental LAC spectra of NV$^-$ centers
    in the SMP2 sample in the magnetic field range 970-1075~G.
    for each curve we give the $f_m$ value used in experiments.
    For the upper curve the phase of the lock-in detector
    is chosen such that the signal for the central LAC-line
    is maximal. For the middle trace the phase is shifted
    by $90^o$ with respect to that for the upper curve. The
    amplitude of the upper curve is increased by a factor of 5,
    for the middle curve -- by a factor of 50. The LAC-lines are indicated by circle, square and
        triangle. (b) Dependence of the amplitude of the three LAC-lines (symbols correspond to the LAC-lines in subplot a) on the modulation     frequency $f_m$. For each curve the magnetic field strength $B_0$ corresponding to the center of the corresponding line is specified.
    For each experimental point the lock-in detector phase is set
    such that the amplitude of the corresponding line was maximal.
    In all cases the modulation amplitude was $B_m=0.5$~G. \label{experiment}}
    \end{figure}

\subsection{LAC-spectra and LAC-lines}

It is well-known that when the sample is oriented precisely,  such
that $[111]||\bm{B}_0$, at a magnetic field $B_0=D=1024$~G there
is a sharp dip, an LAC-line, in the luminescence intensity, which
is caused by an LAC between the $M_S=0$ state and $M_S=-1$ state
of the ground state of the NV$^-$ center. Additionally, weak
satellite lines can be observed; observation of these lines is a
good indication for precise orientation of the sample. Our
experimental results for the SMP2 sample using lock-in detection
are presented in Fig. \ref{experiment}. In Fig.
\ref{experiment}(a) we show the LAC-spectra of the NV$^-$ centers
for the magnetic field range around 1024~G at the modulation
frequency of 12.5~kHz and 17~Hz. When lock-in detection is used,
each LAC-line has two components, a negative one and a positive
one. One can readily see that at $f_m=12.5$~kHz and appropriate
setting of the lock-in detector phase (set such that the amplitude
of the LAC line at 1024~G is maximal) the satellites at 1007~G and
1037~G are hardly visible. These lines are coming from magnetic
dipole-dipole interaction of NV$^-$ centers with neutral nitrogen
(P1 centers), which replaces carbon atoms in the diamond crystal
lattice. The small amplitude of these lines indicates a slow
response times of processes, which lead to the formation of these
lines. For the same reason, the phase shift is close to $90^o$:
when such a phase shift is introduced the satellite lines become
clearly visible. Interestingly, despite the almost ten-fold
decrease of the line intensity (for the central LAC-line) other
weak LAC-lines appear in the spectrum.

When the modulation frequency is reduced to 17~Hz the amplitude
of the central LAC-line increases by roughly a factor of 7,
whereas the satellite lines become 50 times more intense. The
phase shift for all lines becomes negligibly small. Additional
weaker lines show up in the spectrum as the signal-to-noise ratio
significantly increases.

In Fig. \ref{experiment}(b) we present the experimental
dependencies of  the line intensities on the modulation frequency.
In these experiments we measured the total (peak to peak)
amplitude of the two-component LAC-lines; the lock-in detector
phase was set for each line in each experiment such that the
amplitude of the line was maximal. One can readily see that in the
frequency range under study the line intensity changes by roughly
two order of magnitude. Interestingly, the frequency dependence is
not exponential and the slope of the curve increases at low $f_m$
frequencies.

The most interesting experimental fact is that the increase of
the line intensities occurs at low frequencies, namely, at $f_m$
values, which are much smaller than any spin relaxation rates of
the NV$^-$ center. Such a strong dependence can be explained by
polarization transfer from electron spins to nuclear spins having
much longer relaxation times \cite{Anishchik2016}.

\begin{figure}
   \includegraphics[width=0.42\textwidth]{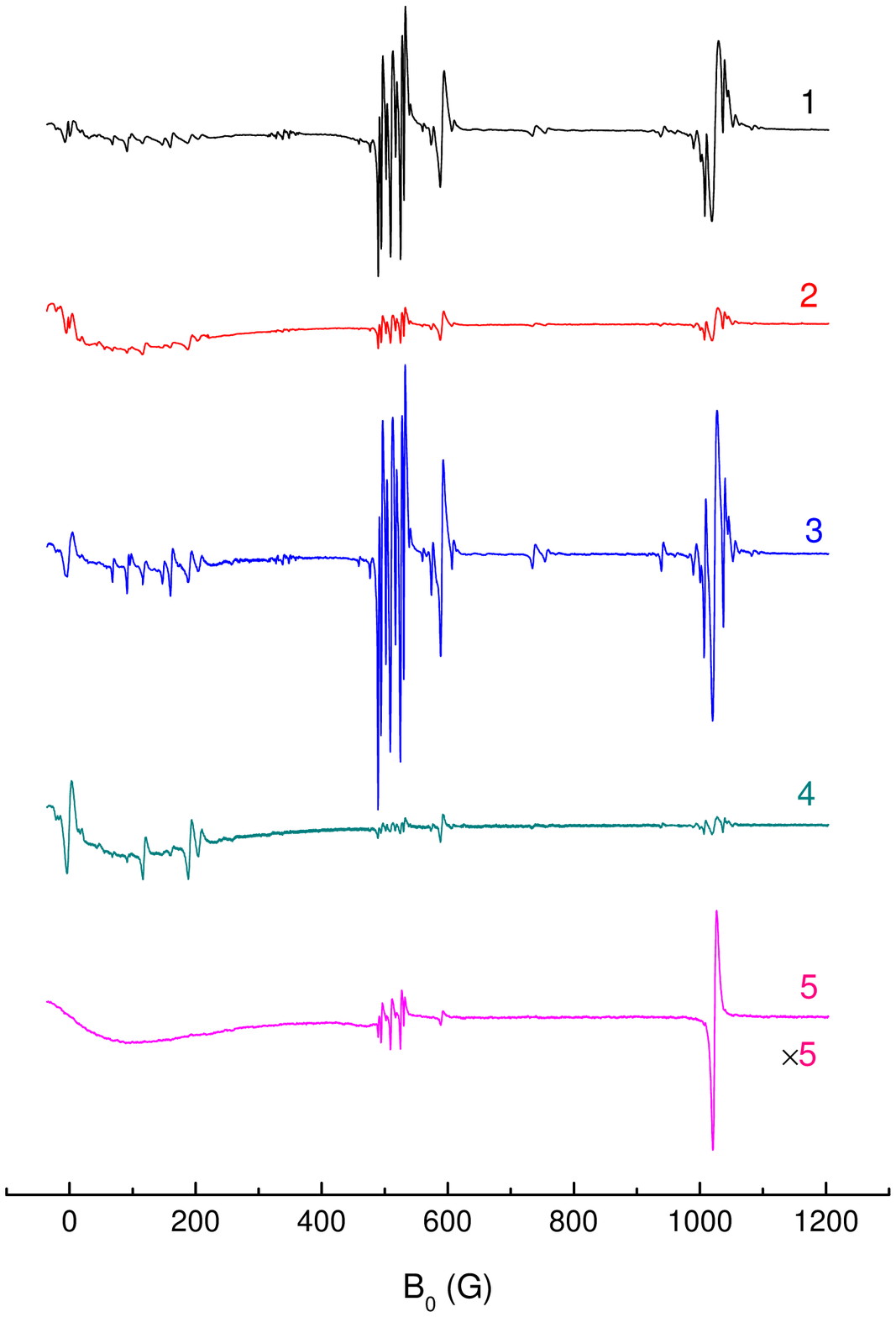}
   \caption{LAC spectra of the SMP1 sample (curves 1, 2) and
   SMP2 sample (curves 3, 4, 5). Experimental parameters: $f_m=17$ Hz
   (curves 1--4) and 12.5 kHz (curve 5).  Polarization of light is
   chosen such that $\bm{E}\bot\bm{B}_0$ (curves 1, 3, 5) and $\bm{E}||\bm{B}_0$
     (curves 2, 4). Curve 5 is multiplied by 5. In all cases
     $[111] || \bm{B}_0$.\label{holeall}}
    \end{figure}

The absence of a phase shift and the significant increase in the
signal-to-noise ratio at low modulation frequencies gives us some
hope to observe new LAC-lines. In Fig.~\ref{holeall} we present
the LAC-spectra of the two samples. These spectra are obtained at
different polarization of the incident light and at $f_m=17$~Hz.
Additionally, we show the LAC-spectrum of the SMP2 sample at
$f_m=12.5$~kHz and $\bm{E}\perp\bm{B}_0$. All previously known
LAC-lines are seen in curve 5 in Fig.~\ref{holeall}, which has
been measured at the high modulation frequency. Namely, these are
the seven lines in the field range 450-550~G, a line at 590 G and
the LAC-line at 1024 G. The zero-field line, which has been found
by some of us \cite{Anishchik2015} is not visible because its
phase shift is about 80$\rm ^o$ at these experimental conditions
resulting in a strongly reduced line intensity.

When the modulation frequency is reduced, many new lines appear
in the LAC spectra of both samples. In the field range 0-250~G
there is a group of lines appearing, which has never been
reported. At the high frequency (12.5~KHz) in this range there is
only one broad line found corresponding to the well-known
\cite{Epstein2005,Rogers2008,Rogers2009,Armstrong2010} smooth
decrease of the photo-luminescence intensity. Besides this,
several groups of low-intensity lines appear. All observed lines
are coming from interaction of the NV$^-$ centers with other
paramagnetic defect centers. Specifically, in the entire spin
system of coupled paramagnetic centers there is polarization
transfer occurring at LACs. As a consequence the luminescence
intensity of NV$^-$ centers is reduced and LAC-lines appear.
Investigating the origin of such LAC-lines allows one to identify
the defect centers and determine their magnetic resonance
parameters, such as zero-field splitting (ZFS) and hyperfine
couplings (HFCs).

When the polarization of the incident light is changed from $
\bm{E}\bot\bm{B}_0$ to $\bm{E}||\bm{B}_0$  the intensity of
LAC-lines  considerably decreases. Specifically, for most lines
the intensity diminishes. However, the intensity of the zero-field
line grows: in spectrum 4 the zero-field line is the most intense.

\begin{figure}
   \includegraphics[width=0.45\textwidth]{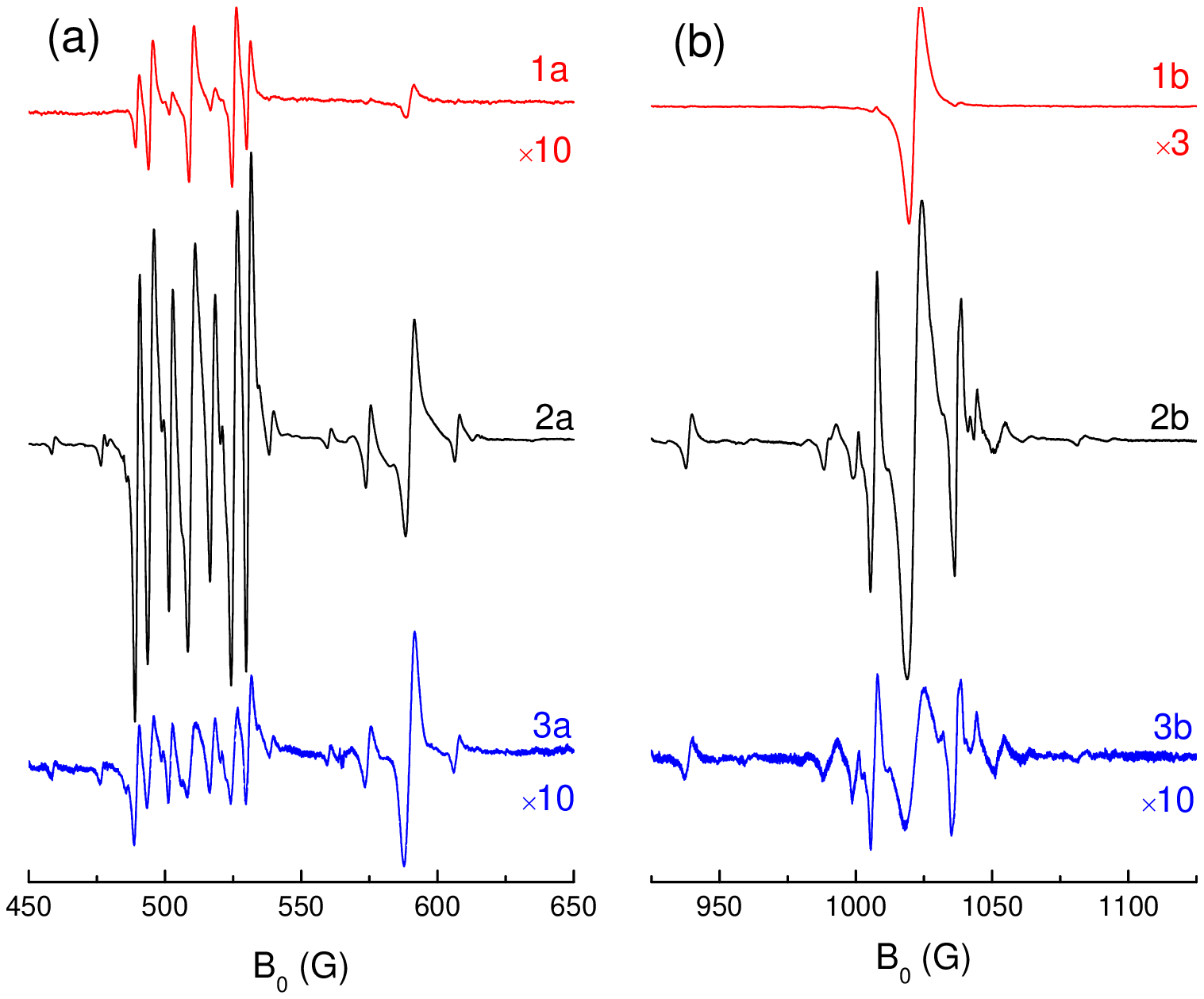} \caption{LAC spectra of the
   SMP2 sample in the range 450--650 G (a) and in the range 925--1125 G (b).
   Experimental parameters:
     $f_m=12.5$ kHz (curves 1a, 1b) and  17 Hz (curves 2a, 2b, 3a, 3b);
      $\bm{E}\perp\bm{B}_0$ (curves 1a, 1b, 2a, 2b), $\bm{E}||\bm{B}_0$
      (curves 3a, 3b). Curves 1a, 3a, 3b are multiplied by 10. Curve 1b
       is multiplied by 3.
      $[111] || \bm{B}_0$.\label{51}}
    \end{figure}

In Fig. \ref{51} we zoom into the LAC-spectrum of the SMP2 sample
showing  the field ranges of 540-650~G and 925-1125~G. Upon
lowering of the modulation frequency from 12.5~kHz to 17~Hz the
intensity of all LAC-lines increases; their relative intensities
also change. Likewise, all lines become much less intense when the
light polarization is $\bm{E}||\bm{B}_0$. For the lines showing up
in the range  580-530 G the intensity decreases by a factor of 40;
for the line at 590~G the decrease is 8-fold. For the LAC-lines
showing up in the range 925-1125~G there is an overall reduction
of the line intensity, which is different for individual lines.

\begin{figure}
   \includegraphics[width=0.35\textwidth]{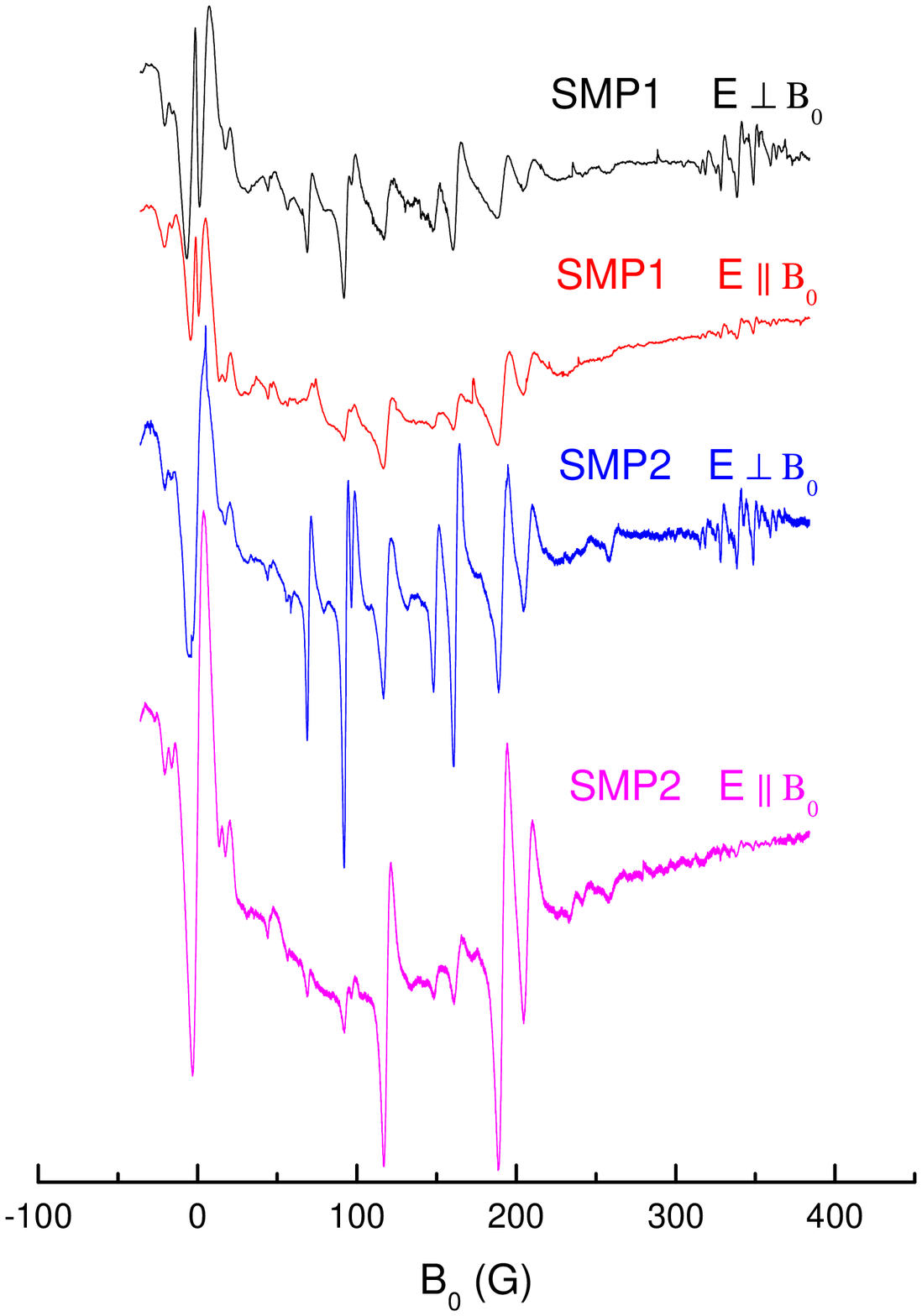} \caption{LAC spectra
   of the SMP1 and SMP2 samples at low fields at different light polarization. \label{leftall}}
    \end{figure}

In Fig. \ref{leftall} we present the LAC spectra for the two
samples  showing the field range from $-40$ to $370$~G at the two
different polarizations of the incident light. For both samples we
can observe sharp LAC-lines, which are much more pronounced for
the SMP2 sample (having a higher concentration of paramagnetic
defect centers). In both samples the zero-field line is observed
\cite{Anishchik2015}; however, its shape is different for the two
samples. The shape and intensity of this line do not change much
upon rotation of the light polarization vector. Additionally,
there is a group of lines appearing in the range 50-250~G; their
intensity is very sensitive to the light polarization. The
spectrum becomes simple for the SMP2 sample when ${\bm E}||{\bm
B}_0$: it  comprises a singlet line at 120~G and a doublet at
about 200~G. These lines are also seen when $\bm{E}\perp\bm{B}_0$
but their intensity a two times lower. At this light polarization,
more lines appear: a singlet at $\sim$50~G, a narrow doublet at
$\sim$100~G and a doublet at $\sim$160~G. At the other
polarization, $\bm{E}||\bm{B}_0$, these lines also show up but
their intensity is so low, that they rather look like experimental
noise.

\begin{figure}
   \includegraphics[width=0.42\textwidth]{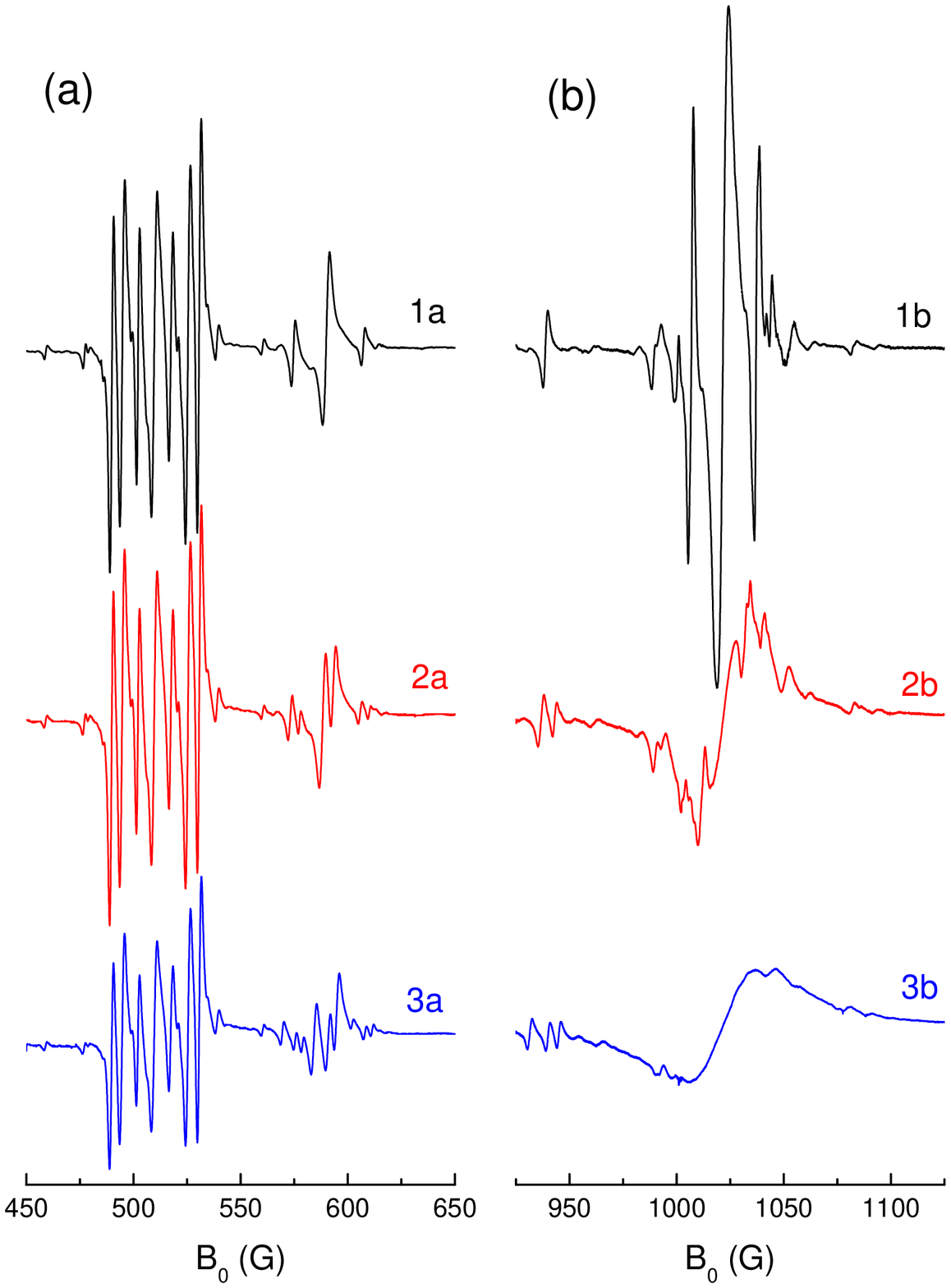} \caption{Variation of
   the LAC spectra of SMP2 upon sample rotation. Here the sample is oriented
   such that the [111] crystallographic axis is parallel to the external
   field $\bm{B}_0$ with a precision of about 0.1$\rm ^o$ (curves 1a and 1b);
   the sample is rotated by 0.4$\rm ^o$ about the [110] axis (curves 2a and 2b);
   the sample is rotated by 1$\rm ^o$ about the [100] axis (curves 3a and 3b).
   Here the modulation frequency is 17~Hz.  \label{turn}}
    \end{figure}

In our experiments the sample is precisely oriented such that
$[111]||{\bm B}_0$.   When the light polarization is
$\bm{E}||\bm{B}_0$ the NV$^-$ centers oriented along the
$\bm{B}_0$ field vector are not excited and do not acquire any
spin polarization. For this reason, the LAC-lines (singlet and
doublet) seen in the lower curve in Fig.~\ref{leftall} are due to
excitation of the NV$^-$ centers tilted by 70.53$\rm ^o$ with
respect to the $\bm{B}_0$ field vector. Other lines, which appear
at the $\bm{E}\perp\bm{B}_0$ polarization are coming from
excitation of the NV$^-$ centers oriented parallel to the
$\bm{B}_0$ field vector.

In Fig.~\ref{turn} we present the dependence of the LAC-line shape
on the sample  rotation, as shown for the field range 450-650 and
925-1125~G. The SMP2 sample is rotated by a small angle. We
compare the results for the sample with the [111] axis oriented
parallel to $\bm{B}_0$ with an accuracy of 0.1$\rm ^o$ and for the
same sample rotated by an angle of 0.4$\rm ^o$ and $\rm\sim 1 ^o$.
The value of the rotation angle is cross-checked by a calculation
performed for the LAC-line at 590~G. One can see that the lines
falling in the range 450-550~G become less intense but keep the
shape and position. The line at 1024~G is broadened and decreased
in amplitude; the satellites practically disappear. The lines at
590~G and 940~G are synchronously split into two or three
components. It is known that the LAC-line at 590~G is due to the
dipole-dipole interaction  of NV$^-$ centers having different
orientations with respect to the field vector
\cite{Armstrong2010}. Most likely, the LAC-line at 940~G has a
similar origin.

\section{Theory}
In order to model the LAC spectra we make use of a model,
developed  previously \cite{Ivanov2008} for simulating coherent
polarization transfer phenomena. In this model we assume that
light irradiation continuously generates electron spin
polarization of the NV$^-$ center. We neglect the spin evolution
occurring during light excitation and relaxation to the ground
state, thereby assuming that such processes are much faster than
spin polarization transfer. Hence, we consider only polarization
from the ground state  of the NV$^-$ centers; polarization
transfer in the excited state can be treated in a similar manner.

When polarization transfer occurs within a single NV$^-$ center it
is  driven by the following Hamiltonian:

\begin{eqnarray}
\nonumber {H}_{NV}
=&\beta\bm{B}_0\bm{g}_1\bm{S}_1+\bm{S}_1\hat{\bm{D}}_1\bm{S}_1+\\
~&+ A_1(\bm{S}_1\cdot\bm{I}_1)+\bm{I}_1\hat{\bm{
Q}}_1\bm{I}_1\label{NVH}
 \end{eqnarray}

Here $g_1$ is the g-factor of the NV$^-$ center,  $\hat{\bm{D}}_1$
is the ZFS tensor of the triplet state, $A_1$ is the HFC constant
with the $^{14}$N nucleus (HFC anisotropy is neglected),
$\hat{\bm{Q}}_1$ is the nuclear quadrupolar interaction, which is
neglected. Hereafter, $\bm{S}_1$ is the spin operator of the
NV$^-$ center (the $S_1$-spin is equal to 1); $\bm{I}_1$ is the
spin operator of the $^{14}$N nucleus of the NV$^-$ center (the
$I_1$-spin is equal to 1). Polarization generated by light
excitation is given by the following density matrix:
 \begin{equation}
    \rho=\rho_{NV}\otimes\rho_{eq}. \label{bmod}
\end{equation}

Here $\rho_{NV}$ is a matrix having only one non-zero element,
which corresponds to the population of the $M_S=0$ state;
$\rho_{eq}$ describes the nuclear density matrix at equilibrium
conditions. In high-temperature approximation this matrix is
simply proportional to the unity matrix of the corresponding
dimensionality, $(2I_1+1)\times(2I_1+1)$, divided by a weighting
factor  in order to provide normalization ${\rm
Tr}\{\rho_{eq}\}=1$. For instance, for a spin-1 nucleus this
matrix has only three diagonal matrix elements equal to $1/3$.

In eq. (\ref{NVH}) we specify the interaction tensors, which
depend on  the sample orientation. These tensors become simple,
namely, diagonal, in the principle axis system. In any other frame
the tensors are no longer diagonal and their precise form are
determined by the frame rotations. The frame rotation can be taken
into account by either performing three Euler rotations or by
specifying the direction cosines for the frame transformations.
Here we do not go into detail of such standard calculations. To
set the orientation of ZFS tensors it is sufficient to perform
only Euler rotation because (i) the experiment is set-up such,
that there is  axial symmetry and (ii) the dipolar tensor is
axially symmetric. We always assume that the $z$-axis is parallel
to the $B_0$ field vector. Thus, to describe the frame
transformation, it is sufficient to specify only the tilt angle,
$\Theta_t$, between the $z$-axis and the $z$-axis of the
$\hat{\bm{D}}$ tensor. Of course, for symmetry reason the latter
$z$-axis coincides with the symmetry axis of the NV$^-$ center;
likewise, all other tensors (g-tensor, HFC tensor) posses the same
symmetry. For NV$^-$ centers the principal values of the ZFS
tensor are:\cite{Doherty2013} $D=2.87$~GHz, $E=0$. In calculations
we always consider NV$^-$ centers oriented parallel to the
external field and NV$^-$ centers tilted by 70.53$\rm ^o$ degrees
with respect to the $z$-axis. In a precisely oriented diamond
crystal a quarter of the NV$^-$ centers have $\Theta_t=0$  and
three quarters have $\Theta_t=70.53\rm ^o$. Rotations of the
sample by small angles can be taken into account in the same way.

To calculate the LAC spectrum we go through the following steps.
First,  we calculate $\rho$ in the eigen- basis of the ${H}_{NV}$
Hamiltonian. Second, we remove all off-diagonal elements
(coherences) of the density matrix thus assuming that they are
washed out during the spin evolution over an extended irradiation
period. Finally, we calculate the luminescence intensity as the
total population of the $M_s=0$ state, $\rho_{00}$, since this is
the state, which provides bright luminescence upon irradiation
when  no spin mixing occurs. Such a method follows closely the
previously developed approach.

By generalizing this approach we can also take into account
polarization  transfer between different paramagnetic defect
centers. For two interacting centers, each having a single
magnetic nucleus, the Hamiltonian takes the form:
\begin{eqnarray}
\nonumber H
=&\beta\bm{B}_0\hat{\bm{g}}_1\bm{S}_1+\bm{S}_1\hat{\bm{
D}}_1\bm{S}_1+A_1
(\bm{S}_1\cdot\bm{I}_1)+\\
 \nonumber  ~&+\beta\bm{B}_0\hat{\bm{g}}_2\bm{S}_2+\bm{S}_2\hat{\bm{D}}_2\bm{S}_2+A_2
 (\bm{S}_2\cdot\bm{I}_2)+\\
 ~&+\bm{S}_1\hat{\bm{D}}_{dd}\bm{S}_2
 \end{eqnarray}
 Here the spin operators and parameters of the second center are introduced
 in the same way (by simply re-defining the indices, $1\to 2$); $D_{dd}\propto
r^{-3}$ (here $\bm{r}_{12}$ is the vector connecting the two
centers)  stands for the tensor of the dipole-dipole interaction
between the electron spins of the two defect centers. The initial
density matrix is constructed in the same way as for a single
NV$^-$ center: the electronic triplet state of the NV$^-$ center is assumed to be
polarized, whereas all other spins are at thermal equilibrium. All
further steps required to calculate the LAC-spectrum are the same
as previously. The method can be generalized to an arbitrary
number of interacting spins in a straightforward way. Finally, to
ease the comparison with the experimental data we can numerically
take the derivative of the calculated LAC-spectra.

Typical magnetic resonance parameters used in calculations
(derived  from experimental data \cite{Doherty2013}) are given in
Table 1.

 \begin{table}\caption{Magnetic resonance parameters used in calculations}
\begin{tabular}{cccccc}
   \hline
\hline
  ~ & D (MHz)& g$_{||}$ & g$_{\perp}$ & ~A$_{||}$ (MHz)~ & ~A$_{\perp}$ (MHz)~ \\
\hline
  NV$^-$ & 2870 & 2.0029 & 2.0031 & 2.5 & 2.5 \\
  P1& 0 & ~2.0023~ & ~2.0023~ & 114 & 81 \\
  NV$^0$ & 1685 & 2.0029 & 2.0035 & 26 & 17.3 \\

     \hline
   \hline
\end{tabular}
  \label{t1}
   \end{table}

\section{Discussion}

\begin{figure}
   \includegraphics[width=0.47\textwidth]{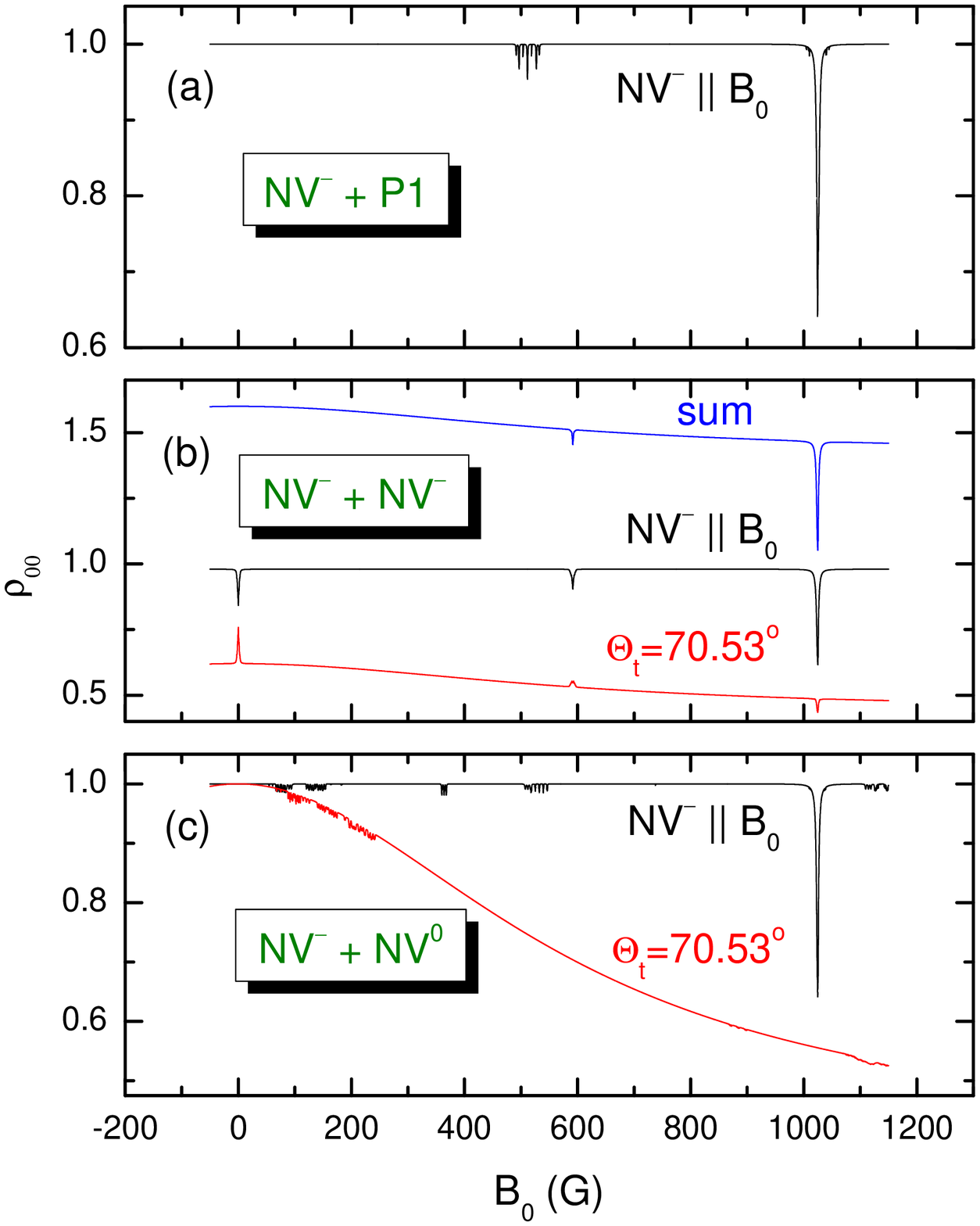}
   \caption{Calculated LAC-spectra for an NV$^-$ center interacting
   with different paramagnetic defect centers. In (a) we show the
   result for and NV$^-$ center interacting with a P1 center; in (b)
    we show the result for two interacting NV$^-$ centers (having
    different orientations and the result of summation over orientations);
     in (c) we show the result for interacting NV$^-$ and NV$^0$ centers.
      \label{theortotal}}
    \end{figure}

In Fig.~\ref{theortotal} we present the calculation results
obtained using  the outlined theoretical model. Specifically, we
present LAC-spectra, which are calculated as the magnetic field
dependencies of $\rho_{00}$. We can now study how $\rho_{00}$ is
modified due to interactions with different paramagnetic defect
centers.

In Fig.~\ref{theortotal}(a) we present the $\rho_{00}(B_0)$
dependence  for an interacting pair of paramagnetic centers
NV$^-$+P1,  where P1 is the neutral nitrogen atom, which replaces
a carbon atom in the crystal lattice. P1 is a paramagnetic
spin-1/2 center; thus, in the calculation we set $S_2=1/2$ and
$I_2=1$ (corresponding to a $^{14}$N nucleus). We also performed
averaging over the four possible orientations of the P1 center.
For simplicity, we show the $\rho_{00}(B_0)$ dependence only for
an NV$^-$ center oriented parallel to the external magnetic field
$\bm{B}_0$. At such an orientation we find sharp dips in the
$\rho_{00}(B_0)$ dependence, which appear at the LACs of the spin
system. One can readily see that in the LAC-spectra there are two
groups of lines. The first group of lines appears around 1024~G
(the main LAC-line and weak satellites); the second group of lines
is around 500~G, altogether there are seven lines in this group.
At other orientations of the NV$^-$ center there are no sharp
lines; for this reason we do not show the $\rho_{00}(B_0)$ for
other orientations.

In Fig.~\ref{theortotal}(b) we also present the calculated LAC
spectra for  two interacting NV$^-$ centers. Here the
$\rho_{00}(B_0)$ dependence is shown for two orientations of the
NV$^-$ centers, both centers are oriented parallel to the external
field and tilted by 70.53$\rm ^o$ with respect to the field. We
also present the sum of the two e $\rho_{00}(B_0)$ dependencies,
which is averaged over four orientations of one of the partners.

At an arbitrary orientation of the NV$^-$ center the LAC-line at
1024~G is  observed. For the NV$^-$ center oriented parallel to
the external field is due to an LAC of the $M_S=0$ and $M_S=-1$
states of the NV$^-$ center itself. When the NV$^-$ center is
tilted by 70.53$\rm ^o$ with resect to the field the LAC-line is
due to polarization transfer between this center and the one
oriented parallel to the field. For this reason, the intensity of
the LAC line in this case is lower.

At both orientations there is an LAC-line at 590~G; however, it
has  different intensity in the two cases. This line is due to
dipole-dipole interaction of two NV$^-$ centers having different
orientations. For the center oriented parallel to the field the
LAC-line is a dip; for the other orientation it is a peak.
However, the amplitude of the LAC-line is higher in the former
case; consequently, the overall effect is negative (a dip is
seen).

Interestingly, at zero field $\rho_{00}$ also has a feature, a
peak  of a dip. In the previous work \cite{Anishchik2015} some of
us have shown that this line comes from interaction of NV$^-$
centers with different orientation. As it is seen from
Fig.~\ref{theortotal}(b), the effect has the same sign but
opposite size for different orientations. Therefore for
$\rho_{00}$ averaged over orientations this line is missing.
However, interpreting experiments where the fluorescence intensity
is measured one should note that the excitation efficiency of an
NV$^-$ center, as well as its spin polarization, are
orientation-dependent. For this reason, the zero-field LAC-line
can be observed; its intensity is expected to be proportional to
the square of the light intensity, as has been found
experimentally \cite{Anishchik2015}.

In Fig.~\ref{theortotal}(c) we present the $\rho_{00}(B_0)$
dependencies  for the case where an NV$^-$ center is coupled to a
neutral NV$^0$ center by dipole-dipole interaction. The electron
spin of the NV$^0$ center is $S_2=3/2$; the nuclear spins are
$I_1=I_2=1$. In the Figure we  show the result for two
orientations of the  NV$^-$, as averaged over the four
orientations of the NV$^0$ center.

\begin{figure}
   \includegraphics[width=0.42\textwidth]{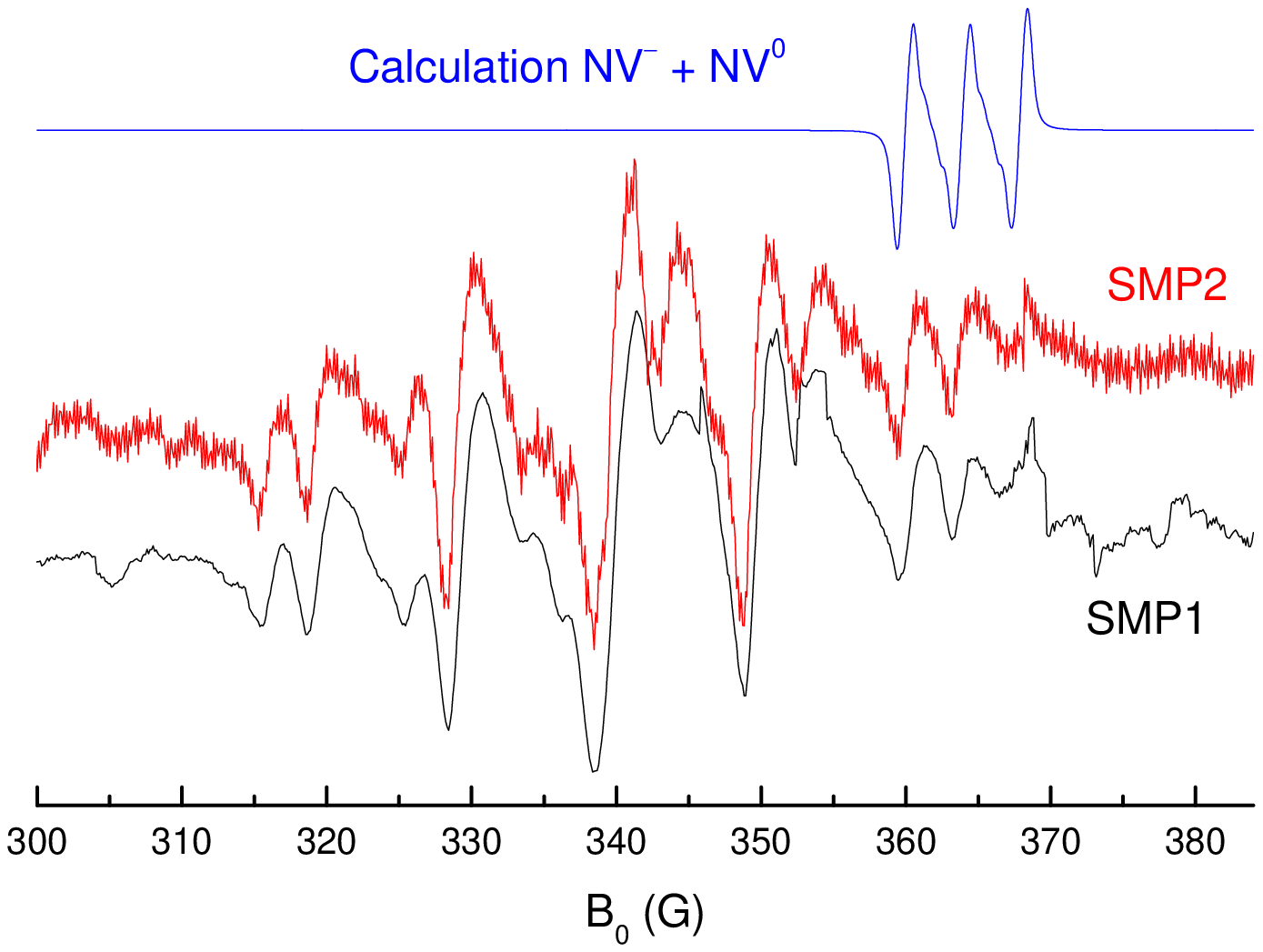} \caption{Experimental LAC-spectra
   for the field range 300-390~G for the two samples obtained at $[111]||\bm{B}_0$
   and $\bm{E}\perp\bm{B}_0$, and the theoretically predicted spectrum for the pair
   NV$^-$+NV$^0$.
    \label{365}}
    \end{figure}

When the NV$^-$ center is oriented along the external field the
LAC-line  at 1024~G is seen as well as additional four groups of
lines. For the tilt angle of 70.53$\rm ^o$ the calculation
predicts numerous lines at magnetic fields less than 300~G. We are
not able to assign any experimentally observed lines except for
the triplet in the field range 360-370~G (see Fig.~\ref{365}).
None of the lines shown in Fig.~\ref{theortotal} can explain the
experimentally  observed lines in the range 25-250~G shown in
Fig.~\ref{leftall}.

For the ease of comparison of theoretical and experimental results
hereafter  we present the first derivative of the $\rho_{00}(B_0)$
curves.

\begin{figure}
   \includegraphics[width=0.4\textwidth]{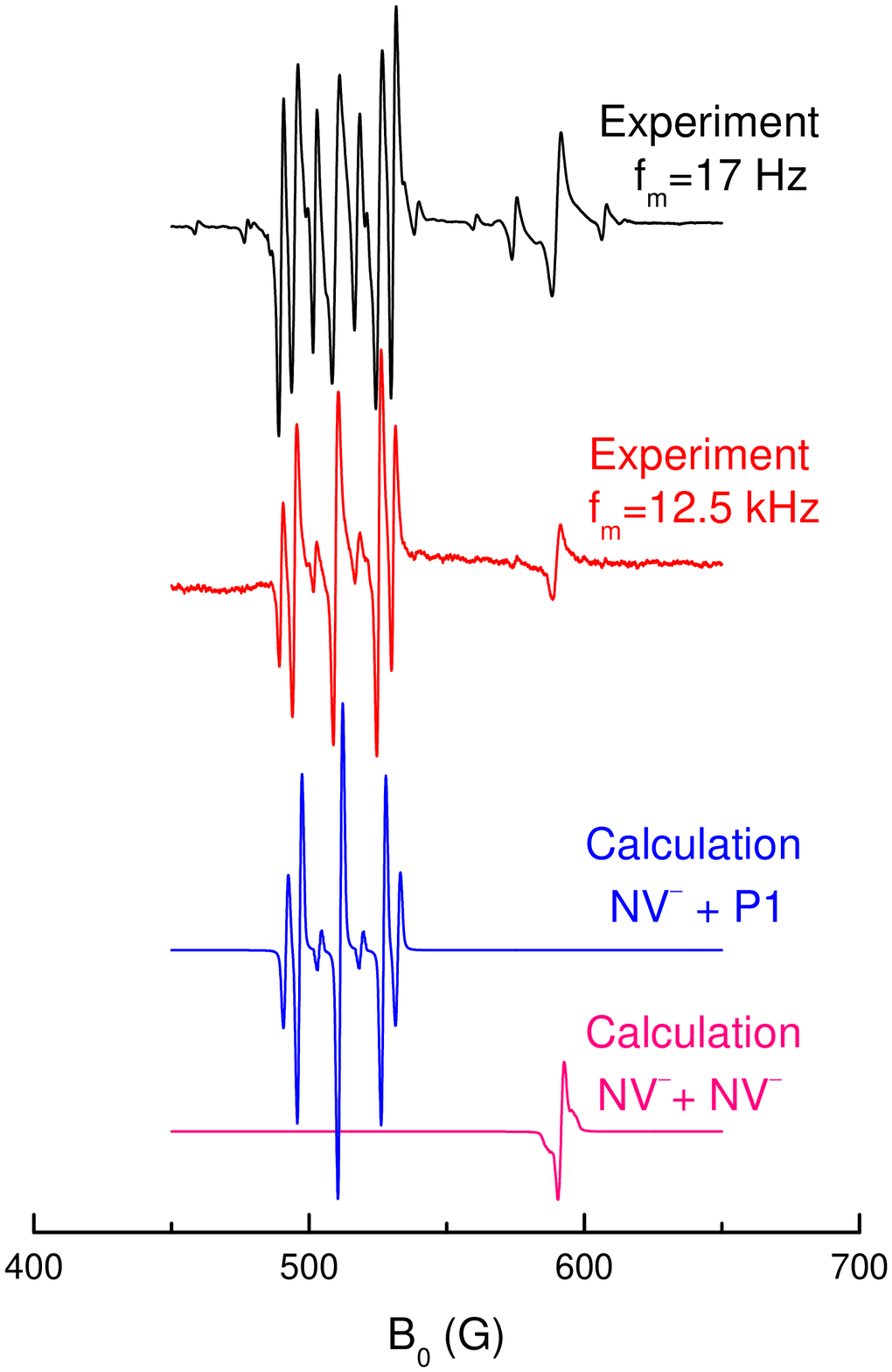} \caption{
   LAC-spectra in the SMP2 sample in the field range 450--650 G at $[111] || \bm{B}_0$ and
   $\bm{E}||\bm{B}_0$. The calculation is performed for interacting pairs of defect centers
   NV$^-$+P1 and NV$^-$+NV$^-$. \label{500et}}
    \end{figure}

\begin{figure}
   \includegraphics[width=0.38\textwidth]{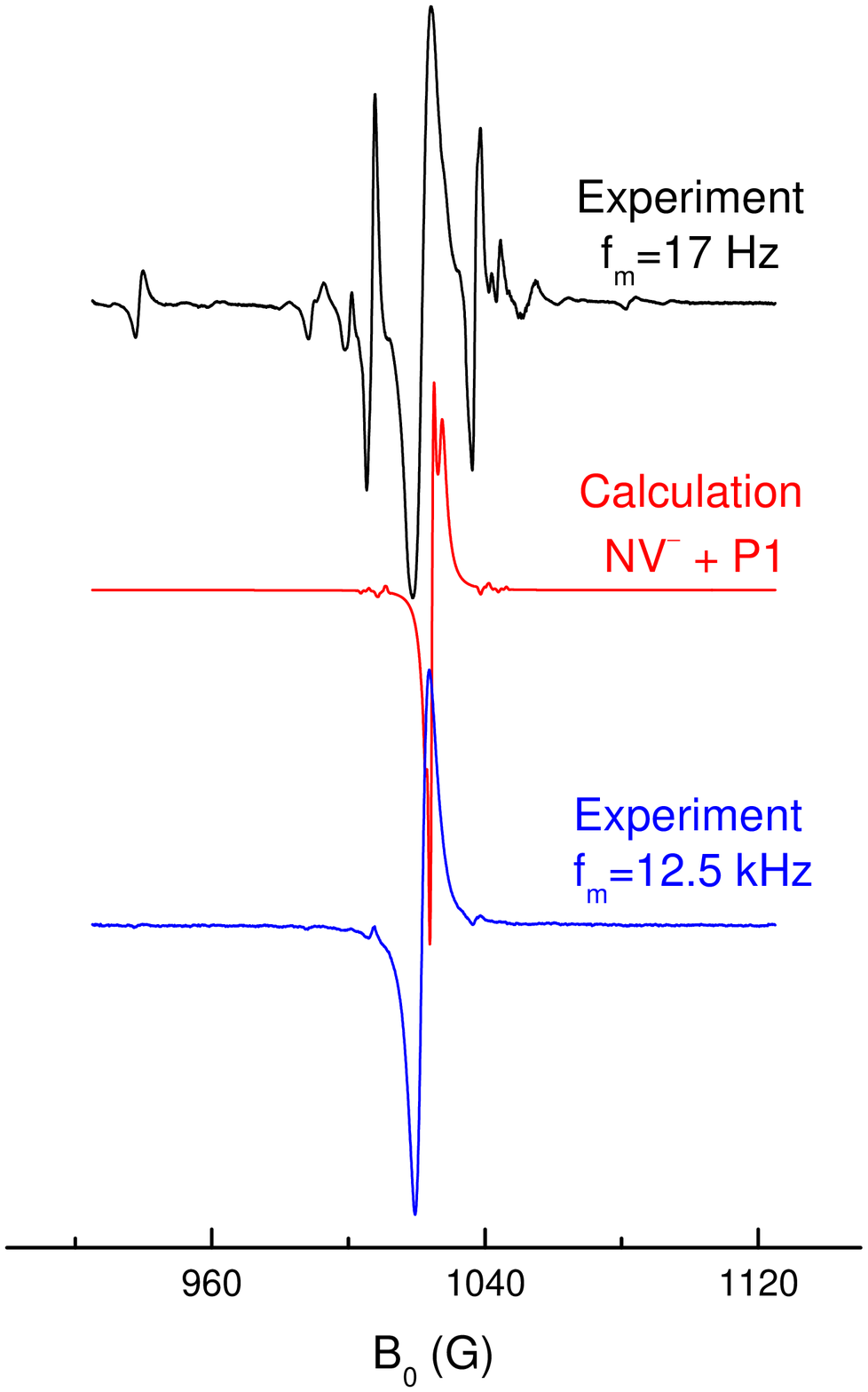} \caption{ Experimental LAC-spectra of the SMP2 sample in the range 925--1125 G and the calculation results for the interacting pair
    NV$^-$+P1. $[111] || \bm{B}_0$ and
   $\bm{E}\perp\bm{B}_0$.\label{1000et}}
    \end{figure}

In Fig.~\ref{500et} two experimental LAC-spectra are shown for the
B$_0$ range  of 450-650~G obtained at the modulation frequency of
17~Hz and 12.5~kHz, as well as the calculation results for the
pairs of defect centers NV$^-$+P1 and NV$^-$+NV$^-$. Interaction
of the NV$^-$ with the  P1 center leads to the formation of
LAC-lines in the field range 480-540~G; their positions exactly
coincide with those for the experimentally measured lines.
Relative intensities of the LAC-lines in the calculation are close
to the experimentally found ones at $f_m=12.5$~kHz being
considerably different from those found at $f_m=$17~Hz. In
Fig.~\ref{500et} we also show the calculation result for the pair
NV$^-$+NV$^-$. The theoretical LAC-line at  590~G is in perfect
agreement with the experiment. However, the satellites, which are
clearly seen at the modulation frequency of 17~Hz, are not
reproduced by the calculation. The most likely reason for the
appearance of these satellites is the splitting due to HFC with
$^{13}$C spins, which are not included in the calculation. As far
as the line at 940~G in Figs.~\ref{turn} and \ref{1000et} is concerned, it can be
reproduced by calculations when we take into account the
interaction of an NV$^-$ center with an excited state of the
NV$^-$ center (although the fitting HFC constant for the excited state turns out to be
significantly smaller than the value found previously
\cite{Steiner2010}). Therefore the origin of this line remains an open issue.

In Fig.~\ref{1000et} we show the experimental LAC-spectra around
the LAC-line at  1024~G obtained using $f_m=17$~Hz and
$f_m=12.5$~kHz and the calculated spectrum for an NV$^-$ center
interacting with a neutral nitrogen P1 center. One can readily see
that (as well as in the previous case) the relation of the
amplitudes of the LAC-lines is close to that found experimentally
using the modulation frequency of 12.5~kHz. However, the
discrepancy with the experimental results obtained using
$f_m=17$~Hz is bigger. Most likely, such a discrepancy comes from
the simplifying assumptions made in the calculations.
Specifically, the calculation reflects the static value of
$\rho_{00}(B_0)$ and the experiment at high modulation frequencies
much better corresponds to the derivative of this static value. At
the same time, the effect of the increased intensity is much more
pronounced for weak LAC-lines, see Fig.~\ref{experiment}. Hence,
we can state that lowering of $f_m$ leads to distortions of
LAC-lines but at the same time allows one to detect weak
LAC-lines.

\begin{figure}
   \includegraphics[width=0.4\textwidth]{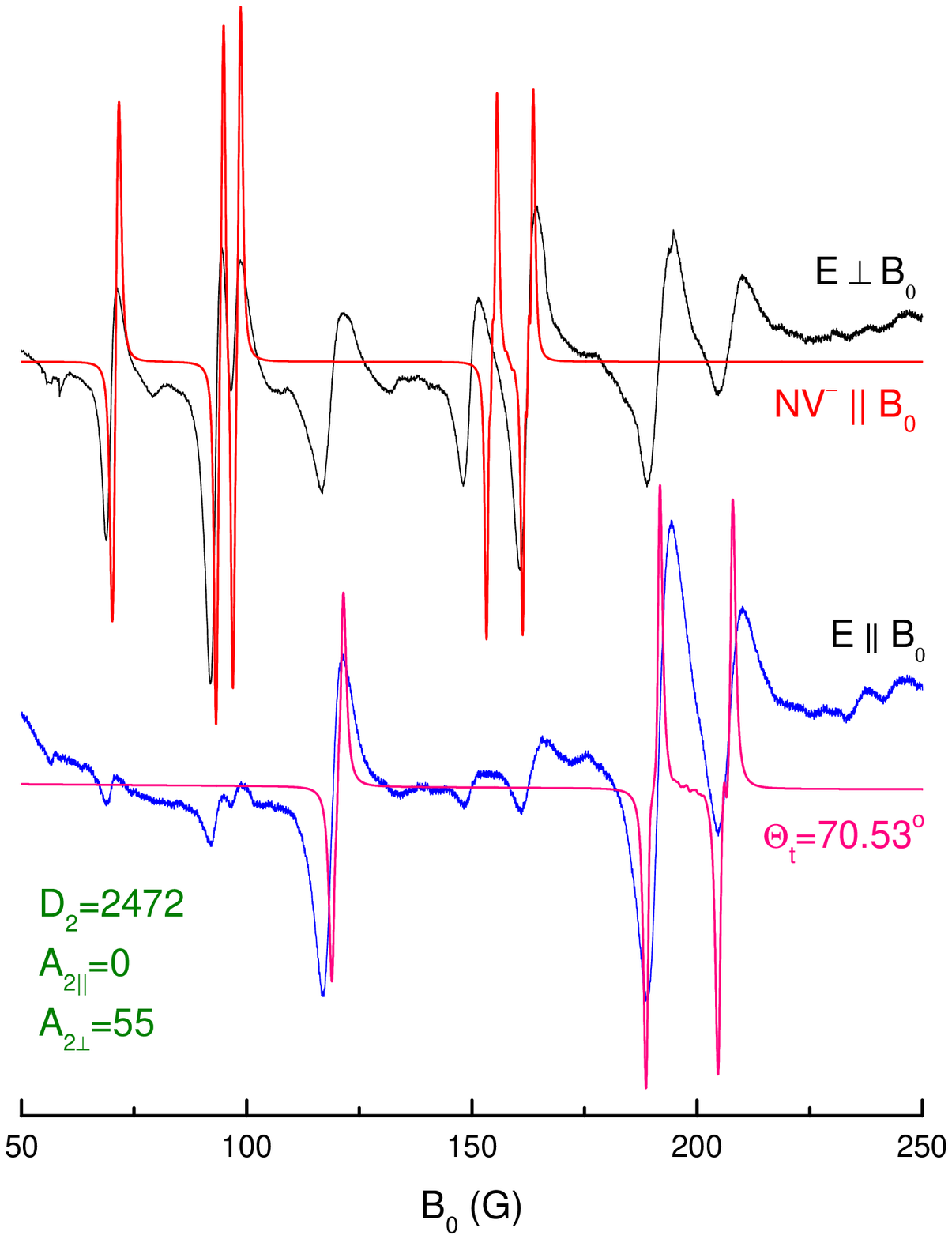} \caption{Simulation
   of LAC-spectra at low fields. Experimental curves are shown for
   $\bm{E}\perp\bm{B}_0$ and $\bm{E}||\bm{B}_0$.  $[111] || \bm{B}_0$;
   here $f_m=17$~Hz. The theoretical curves are obtained for
     an NV$^-$ center interacting with a paramagnetic center with $S_2=1$
     and $I_2=1/2$ for the two orientations of the NV$^-$
center. The fitting parameters are given in the Figure.
   \label{leftdiff}}
    \end{figure}

The experimental results for the fields lower than 250~G cannot be
explained by considering interaction of an NV$^-$ center with the
paramagnetic centers (P1, NV$^-$ and NV$^0$) mentioned above. We
tried to determine the parameters, which provide good agreement
between theory and experiments. It turns out that the experimental
curves can be reasonably well modeled by considering the
dipole-dipole interaction with a paramagnetic center with the
electron spin $S_2=1$ and nuclear spin $I_2=1/2$, having the same
symmetry as the NV$^-$ center. In Fig.~\ref{leftdiff} we show the
experimental traces and calculation results. We did not attempt to
fit the width of the LAC-lines and fitted only the positions of
the lines in the LAC-spectra. The experimental data obtained for
$\bm{E}\perp\bm{B}_0$ are compared to the calculation assuming
that the NV$^-$ center axis is parallel to the magnetic field
(since the centers with this orientations are most effectively
excited and polarized by light). When $\bm{E}||\bm{B}_0$ the
NV$^-$ centers, whose axis is parallel to $\bm{B}_0$, are not
excited at all. For this reason, we compare the experimental
spectra with the calculations assuming that the NV$^-$ center is
tilted by 70.53$\rm ^o$ with respect to $\bm{B}_0$. As can be seen
from the Figure one can fit the positions of almost all LAC-lines
by using the following parameters: $D_2$=2472~MHz, $E_2$=0,
$A_{2||}$=0, $A_{2\perp}$=55~MHz.The only line, which cannot be
assigned is the line at 150~G. Imperfections of the fit and
somewhat unusual relation between the components of the HFC tensor
leaves us in doubts about the model. Most likely, the unknown
center has a different symmetry. We cannot identify what this
defect center is; one can only assume that this is a center
containing a substitutional phosphorus atom and a vacancy next to
it.

Finally, let us discuss the lines found in the range 300-390~G,
see Fig.~\ref{365}. One can see that there are many lines in this
range.  The calculation for the pair NV$^-$+NV$^0$ reproduces well
three of these lines. The origin of other lines remains unclear.

Additional calculations validating the theoretical model  and
showing the optimal fitting parameters are shown in Supplementary
Information at the end of the manuscript.

\section{Conclusions}
In this work by using lock-in detection we have found many new
LAC-lines  in the magnetic field dependence of the luminescence
intensity of the NV$^-$ centers in diamonds. These lines strongly
depend on the modulation frequency: by lowering $f_m$ one can
increase the line intensity and resolve new LAC-lines. We have
found a simple and efficient method for describing spin
polarization transfer between the NV$^-$ centers and other defect
centers in the diamond crystal. This method allows one to
calculate LAC-spectra and assign LAC-lines. We were able to
identify a previously unknown defect center. Potentially, our
experimental method going hand in hand with the new theory is a
powerful tool for investigating paramagnetic defect centers and
their interaction.

\begin{acknowledgments}

Experimental work was supported by the Russian Foundation for
Basic Research (Grant No. 16-03-00672); theoretical work was
supported by the Russian Science Foundation (grant No.
15-13-20035).

\end{acknowledgments}

\bibliography{LAC2}

\begin{widetext}
~
\end{widetext}

\begin{widetext}

\section*{Supplementary information}

In Figs. \ref{leftD2}--\ref{leftAper} we present the experimental
LAC-spectra of the SMP2 sample (for different light polarizations)
along with theoretical simulations at different parameters. These
figures clarify the relevant parameters as extracted from fitting.
We shown only the field range from $-50$ to $400$ G. For the
unknown paramagnetic center, X, we assume that its nuclear spin is
equal to 1 and $I_2=1/2$. In all figures in the left we show the
results for the NV$^-$ centers oriented along the magnetic field;
in the right - for the centers tilted by 70.53$\rm ^o$ with
respect to  $\bm{B}_0$. In all cases we assumed that the X-center
has the same symmetry as the NV$^-$ center and averaged over its
four possible orientations.

In Fig. \ref{leftD2} we show the calculation performed assuming
$A_2=0$ with different $D_2$ values. The $E$ parameter is equal to
zero in all cases; the $D_2$ value (in MAX) corresponds to the
same color of the curve. One can see that the best agreement with
the experimental results is reached when $D_2=2475$~MHz.

In Fig. \ref{leftAinv} we assume that the HFC of the X-center  is
isotropic and vary its value. It is seen the experimental
LAC-spectrum cannot be simulated at any non-zero HFC value.

In Fig. \ref{leftApar} we assume that the HFC censor is strongly
anisotropic, namely,  $A_{2\perp}$=0. Clearly the quality of the
fits is the same as in the calculation assuming isotropic HFC.

Finally in Fig. \ref{leftAper} we assume that the HFC tensor is
anisotropic but set $A_{2||}$=0. The agreement with the experiment
becomes much better; however, the reason for $A_{2||}$ is unclear.

\begin{figure}
   \includegraphics[width=0.87\textwidth]{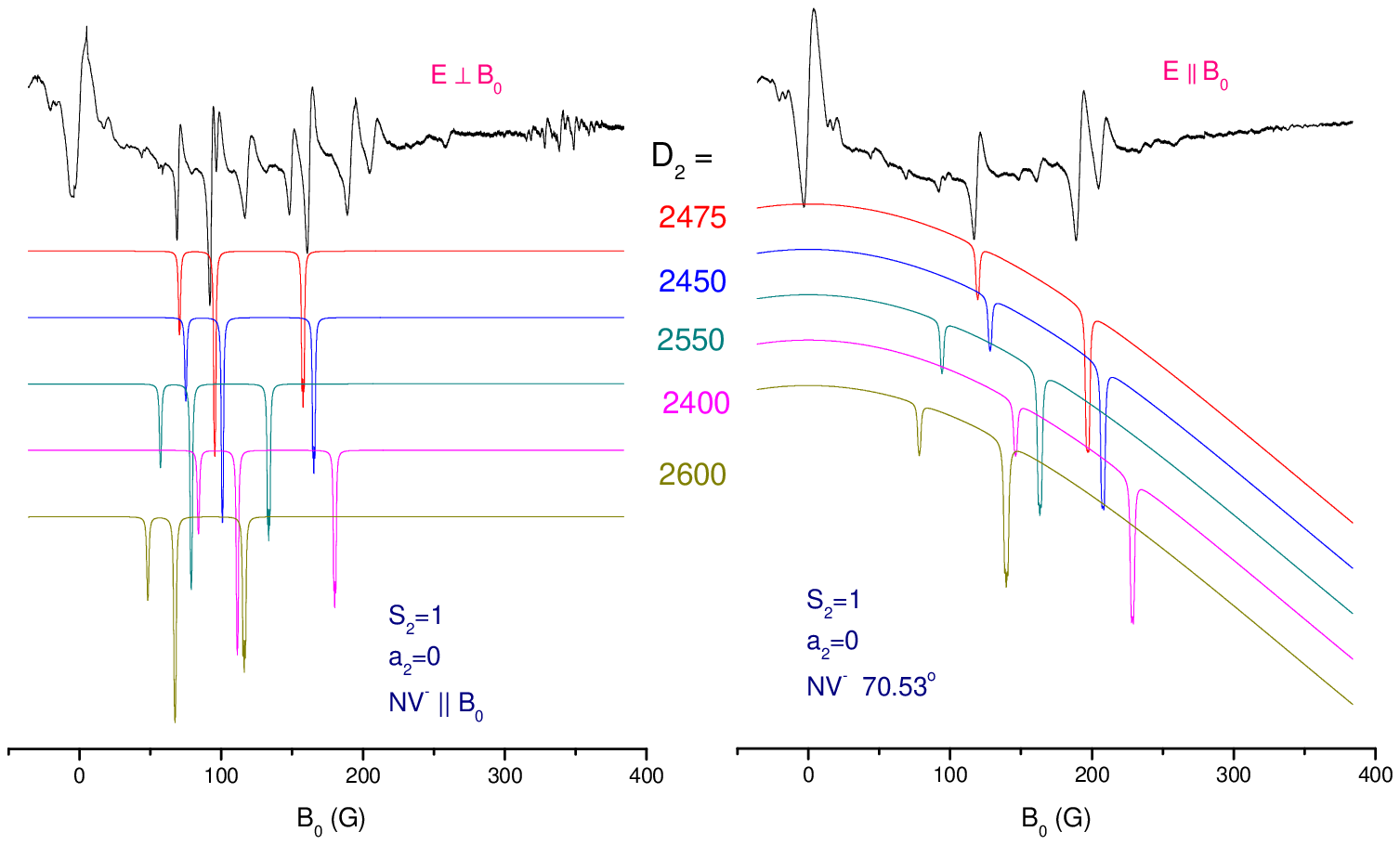} \caption{The
    dependence of the LAC-spectra on the $D_2$ parameter. \label{leftD2}}
    \end{figure}

\begin{figure}
   \includegraphics[width=0.87\textwidth]{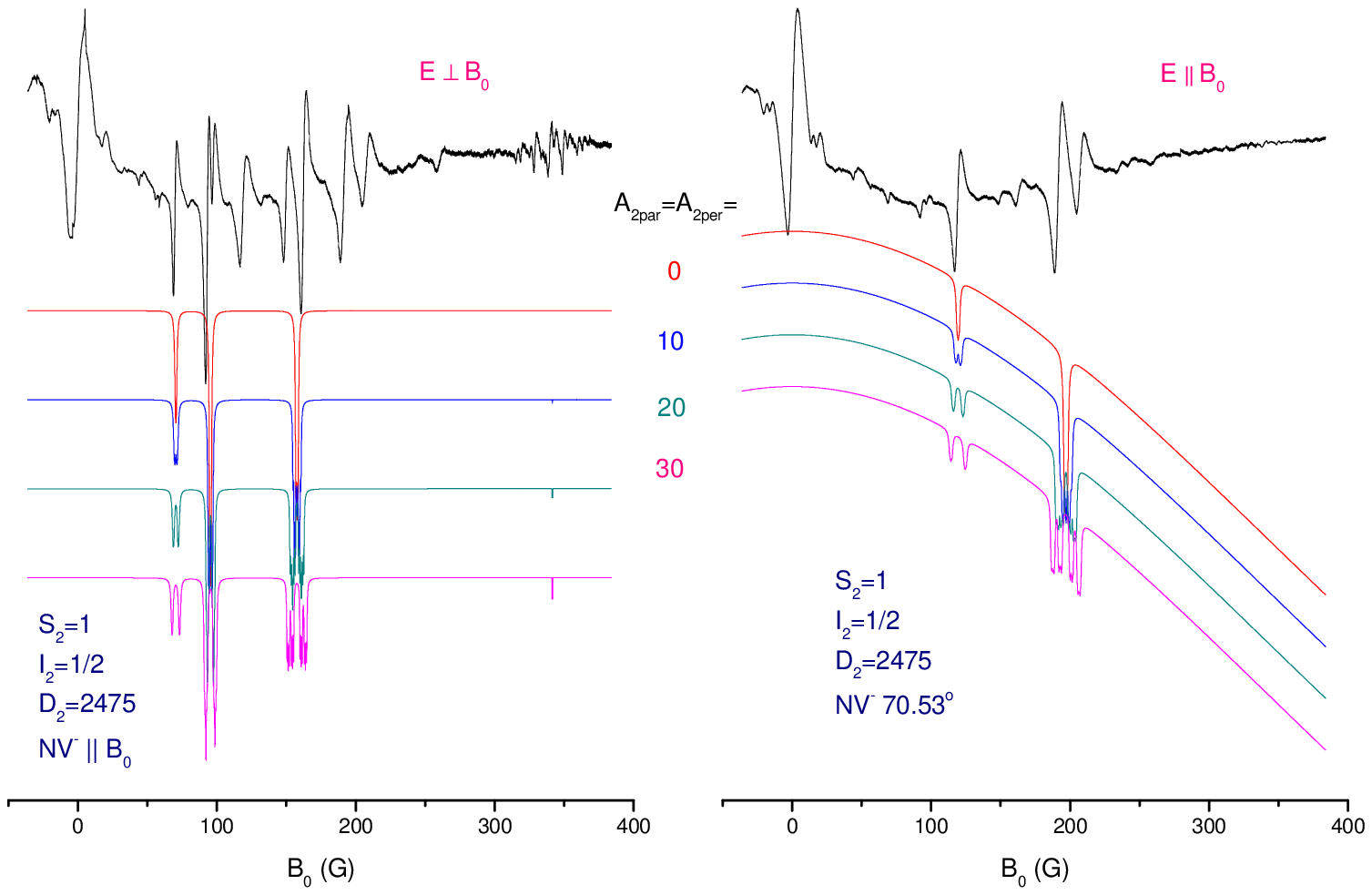} \caption{The
    dependence of the LAC-spectra on HFC, assuming isotropic HFC. \label{leftAinv}}
    \end{figure}

\begin{figure}
   \includegraphics[width=0.8\textwidth]{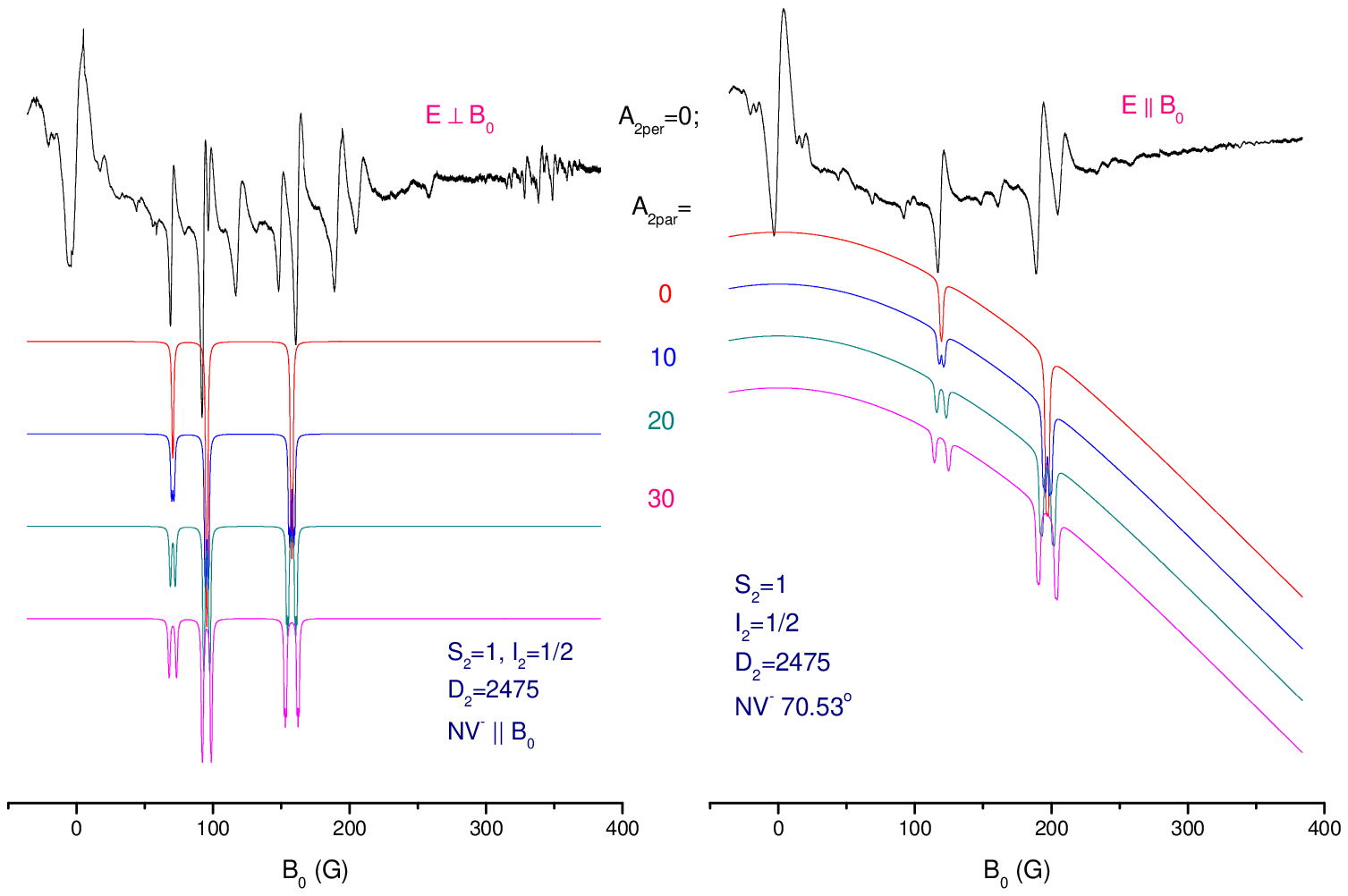} \caption{Dependence
    of the LAC-spectra on HFC assuming $A_{2\perp}$=0).
   \label{leftApar}}
    \end{figure}

\begin{figure}
   \includegraphics[width=0.8\textwidth]{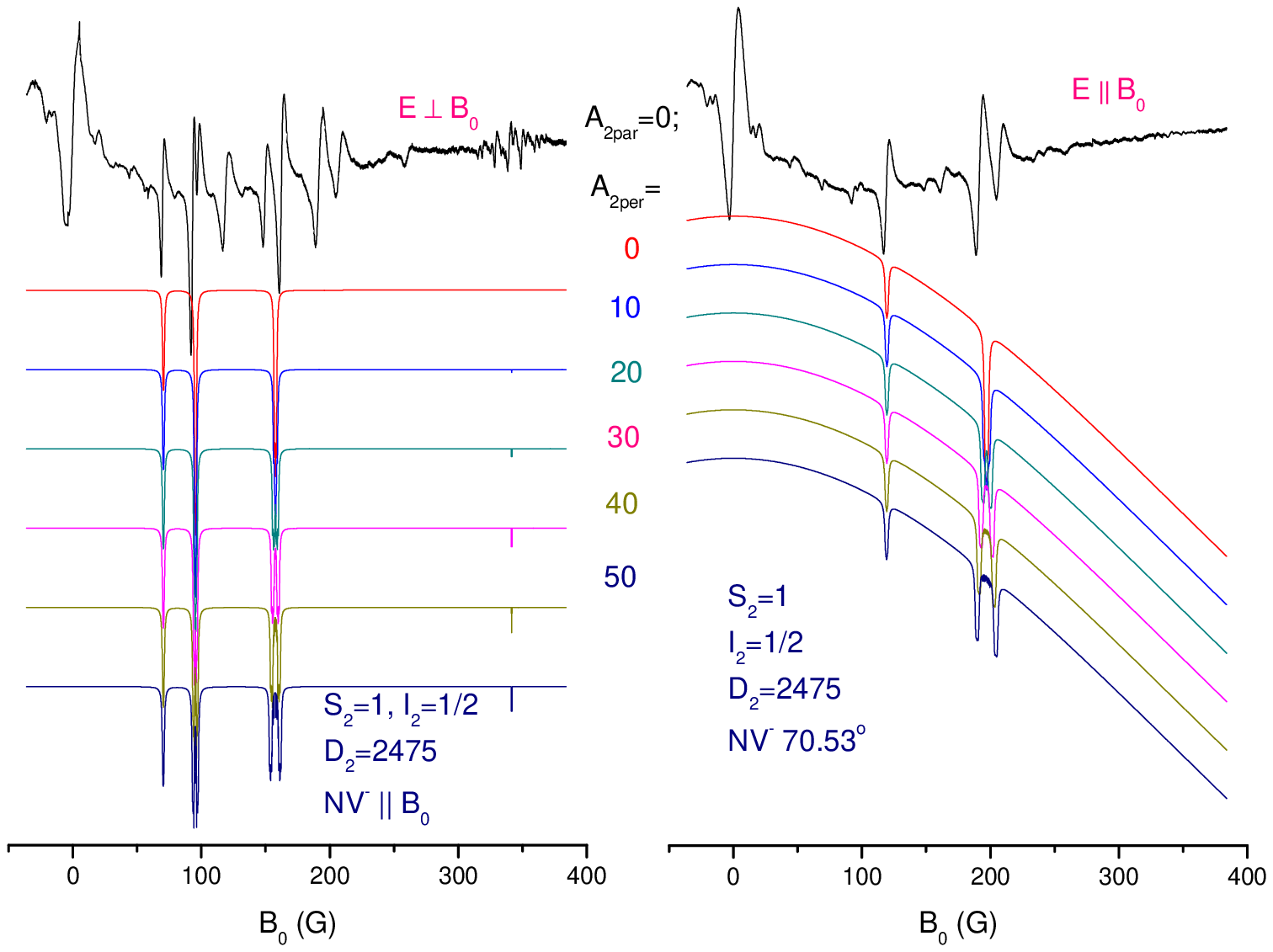} \caption{Dependence
    of the LAC-spectra on HFC assuming $A_{2||}$=0). \label{leftAper}}
    \end{figure}

\end{widetext}

\end{document}